\documentclass[%
reprint,
superscriptaddress,
showpacs,preprintnumbers,
 amsmath,amssymb,
 aps,prl
]{revtex4-1}

\usepackage{graphicx}
\usepackage{dcolumn}
\usepackage{bm}
\usepackage[utf8x]{inputenc}
\usepackage{color}
\usepackage{soul}

\begin{document}

\title{Deviations from the extended London model at high magnetic fields in YBa$_2$Cu$_3$O$_7$}

\author{E. Campillo}
\email{emma.campillo@sljus.lu.se}
\affiliation{Division of Synchrotron Radiation Research, Lund University, SE-22100 Lund, Sweden}

\author{M. Bartkowiak}%
\affiliation{%
 Helmholtz-Zentrum Berlin f\"ur Materialien und Energie,Hahn-Meitner-Platz 1, D-14109 Berlin, Germany
}%

\author{R. Riyat}
\author{E. Jellyman}
\affiliation{School of Physics and Astronomy, University of Birmingham, Birmingham B15 2TT, United Kingdom}

\author{A. S. Cameron}
\affiliation{
 Institut f\"ur Festk\"orperphysik und Materialphysik, Technische Universit\"at Dresden, D-01069 Dresden, Germany
}%

\author{A. T. Holmes}
\affiliation{%
 European Spallation Source ERIC, P.O. Box 176, SE-221 00, Lund, Sweden}%

\author{O. Prokhnenko}
\author{W.-D. Stein}%
\affiliation{%
 Helmholtz-Zentrum Berlin f\"ur Materialien und Energie,Hahn-Meitner-Platz 1, D-14109 Berlin, Germany
}%

\author{A. Erb}
\affiliation{%
 Walther Meissner Institut, BAdW, D-85748 Garching, Germany
}%

\author{E. M. Forgan}%
\affiliation{School of Physics and Astronomy, University of Birmingham, Birmingham B15 2TT, United Kingdom}

\author{E. Blackburn}%
 
 \affiliation{Division of Synchrotron Radiation Research, Lund University, SE-22100 Lund, Sweden}

\date{\today}

\begin{abstract}
We report on the evolution with magnetic field and temperature of the vortex lattice (VL) in fully-oxygenated YBa$_2$Cu$_3$O$_7$ as studied by time-of-flight small-angle neutron scattering. Using the HFM/EXED beamline, we have obtained data up to 25.9 T - much higher than that available previously. Our VL structure results indicate the progressive suppression by field of the superconductivity along the crystallographic $\bf{b}$ (CuO chain) direction. The intensity of the diffracted signal reveals the spatial variation of magnetisation caused by the VL (the ``form factor''). Instead of a rapid fall-off with field, as seen in superconductors with smaller upper critical fields, we find that the form factor is almost constant with field above $\sim$ 12 T. We speculate that this is due to Pauli paramagnetic moments, which increase at high fields due to alignment of the spins of quasiparticles in the vortex cores.

\end{abstract}

\maketitle



\section{Introduction}

The vortex lattice (VL) in YBa$_2$Cu$_3$O$_{7-\delta}$ (YBCO) has been studied by small-angle neutron scattering (SANS) for over a quarter of a century, with the first observation made in a magnetic field of just 0.2 T \cite{For90}. Enormous advances in sample quality and SANS sample environments have allowed VL studies in YBCO to flourish and yield much information about superconductivity in this material~\cite{Yet93a,Yet93b,Kei94,Kei93,For95,Joh99,Bro04,Sim04,Whi08,Whi09,White2011,Cam14}. In this paper, we present data on fully-oxygenated YBCO in fields up to 25.9~T -  experimental results obtained from the HFM/EXED neutron beamline at the Helmholtz Zentrum Berlin~\cite{Pro17,Sme16,Pro15}. SANS measurements of the VL structure as a function of field and temperature give us information on the penetration depth, coherence length, and the superconducting gap structure of a given superconductor. In YBCO, SANS can also reveal the effective mass anisotropy, VL melting, VL pinning \cite{Cam14}, and field-induced non-locality \cite{White2011}.

YBCO has an orthorhombic crystal structure, with CuO chains running along the {\bf b} direction as well as the nearly-square CuO$_2$ $ab$ planes which are common to all cuprate superconductors. On cooling from the disordered high-temperature tetragonal phase, YBCO naturally becomes twinned with \{110\} twin boundaries separating domains with interchanged  {\bf a}- and  {\bf b}-axes. These twin planes act as strong pinning centers and control the VL orientation or structure observed in early SANS studies on twinned single crystals of YBCO \cite{For90,Yet93a,Yet93b,Kei94,Kei93,For95,Bro04,Sim04,Whi08}. Rotation of the twin boundaries out of the field direction was somewhat effective in suppressing the pinning effects on the VL \cite{Yet93a,Bro04}. However, a more effective way to reduce the effects of the twin boundaries is to make measurements on a de-twinned sample. In the first study to do so \cite{Joh99}, the VL diffraction pattern showed a two-fold symmetry, which reflected the $a$-$b$ anisotropy of superconductivity in YBCO. The sign of this anisotropy showed that carriers in the CuO chains contributed to the superfluid density along the  {\bf b} direction. 

The availability of high-quality de-twinned single crystals of YBCO and high magnetic fields on SANS beam-lines has allowed further study of this intrinsic VL structure, which shows that the field-induced VL structural transitions are \emph{first-order}, unlike the \emph{second-order} structural transitions observed in the twinned samples \cite{Bro04,Whi08}. Here we extend the field range up to 25.9~T, obtaining new information on the intrinsic VL structure and superconducting state at high fields, including the temperature dependence at 25.9 T.  

\section{Experimental details}

The fully-oxygenated sample was a mosaic of aligned single crystals with total mass $\sim 70$~mg and an (overdoped) $T_{\rm c}$ of $\sim 89$~K. It is further described in the Appendix. It was mounted with the crystal \textbf{c} axis parallel to the horizontal applied field and the \textbf{a} axis horizontal. The in-plane orientation of the mosaic is rotated by $90^{\circ}$ with respect to that in previous work performed on the same sample at lower fields \cite{White2011,Cam14}.

Our neutron measurements were carried out in two different experiments at the High Magnetic Field Facility for Neutron Scattering \cite{Pro17} which consisted of the High Field Magnet (HFM) \cite{Sme16} and the EXtreme Environment Diffractometer (EXED) \cite{Pro15} at the Helmholtz-Zentrum Berlin (HZB). The HFM was a hybrid solenoid magnet system with a maximum field of 25.9~T, making it the highest continuous magnetic field available in the world for neutron scattering experiments at the time. The direction of the horizontal magnetic field, and therefore of the sample, could be rotated relative to the incoming beam by up to 12$^{\circ}$, limited by the size of the conical solenoid openings. The multi-purpose HFM/EXED instrument operated in time-of-flight (TOF) mode, with a wide range of incident neutron wavelengths, maximising the volume of reciprocal space that can be observed for a given orientation of the HFM. In our experiments, we chose this range to be 2.55-8.15 \AA~for the first experiment and 2.3-9 \AA~for the second one. Our data resulted from the first use of this facility in SANS mode, enabling neutron measurements of mesoscopic magnetic structures in high fields.

The VL was prepared for observation at the base temperature of 3~K, by cooling the sample through $T_{\rm c}$ in an applied magnetic field. The VL quality is usually improved by oscillating the field value while cooling \cite{Cam14}. In the present case, the small variations $\sim 30$ mT from the magnet power supply served this purpose. For a given value of applied field, and given rotation of the HFM away from the incident beam direction, only one particular wavelength of neutron would be incident at the Bragg angle for diffraction by the VL. Neutrons of different wavelengths in the range supplied in TOF mode would be incident at angles away from the Bragg condition. Hence the data at a single sample angle can contain a substantial part of the `rocking curve' of intensity of the VL Bragg spot. This contrasts with experiments using a monochromatic neutron beam, where the integrated intensity under the `rocking curve' is obtained by taking measurements at many different sample angles, rocking through the Bragg condition. In this work, TOF measurements were taken at just a few sample angles to check for consistency and to ensure that the entire `rocking curve' is covered by the wavelength spread \cite{Pau12}. The VL diffraction pattern shown in Fig. 1 was obtained by measuring at 3~K for both positive and negative magnet rotation angles, to give a complete 1st-order diffraction pattern from the VL. Background measurements were taken at the same angles above $T_{\rm c}$ and were subtracted from the measurements below $T_{\rm c}$ so that only the VL signal remained.

Data visualization and  analysis were performed using the Mantid software package \cite{Arn14}. This allowed us to determine both the VL structure and the magnitude of the spatial variation of induction within the VL as a function of applied magnetic field.

\section{Results}

\begin{figure}
\includegraphics[width=\columnwidth]{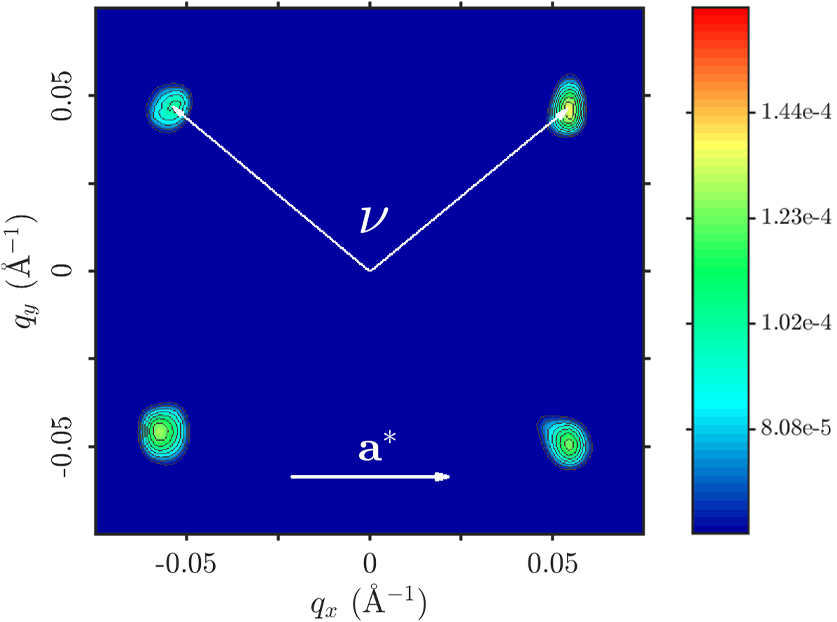}
\caption{A vortex lattice diffraction pattern at 23 T and 3 K. The opening angle, $\nu$, is used to describe the structure of the vortex lattice. The plotted signal is a measure of the counts per pixel summed along $q_z$ divided by the product of the incident beam intensity and the square of the neutron wavelength.} 
\end{figure}


The VL structure can be described by the angle between two diffraction spots, $\nu$, which is bisected by the $\bf{b}$* direction. Fig.~1 shows a typical diffraction pattern at 23~T and 3~K. The incident beam has been masked out, and the data were smoothed using Mantid.

Fig.~2(a) shows the opening angle, $\nu$, as a function of magnetic field. The circular points represent measurements from this study, and for comparison we have included data for the opening angle from previous measurements \cite{White2011,Cam14}. We see that the structure evolves continuously through a square VL at approximately 11.5~T and $\nu$ increases up to approximately 100$^\circ$ at the maximum applied field of 25~T. All measurements were taken at 3 K.
In Fig.~2(b) the temperature dependence of the opening angle, $\nu$, is displayed. At 25 T and 25.9 T there are no apparent changes as a function of temperature in the opening angle which remains around $100^{\circ}$.  The value at 19~T lies close to $95^{\circ}$, as expected from Fig.~2(a).

\begin{figure}
\includegraphics[width=\columnwidth]{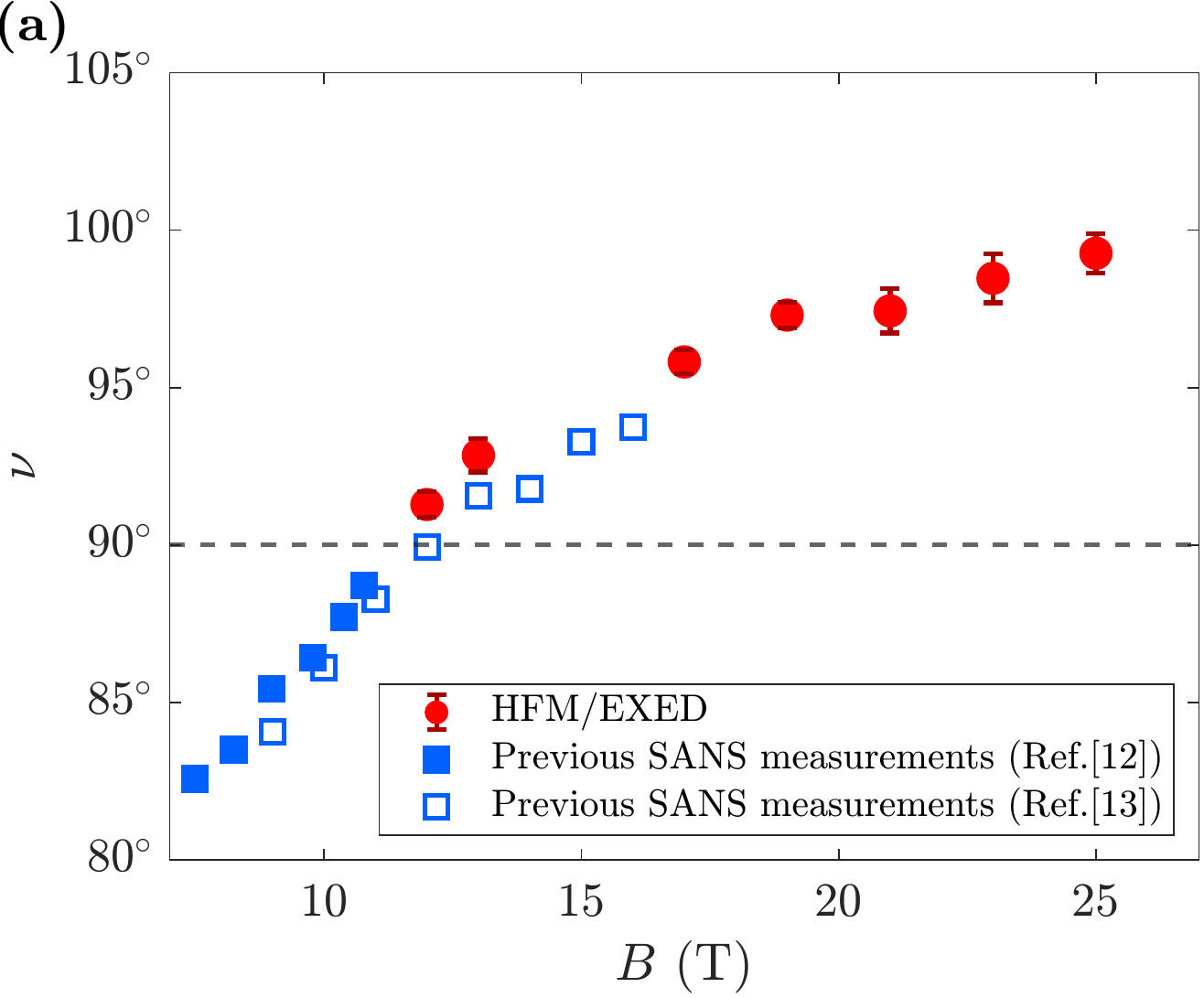}
\includegraphics[width=\columnwidth]{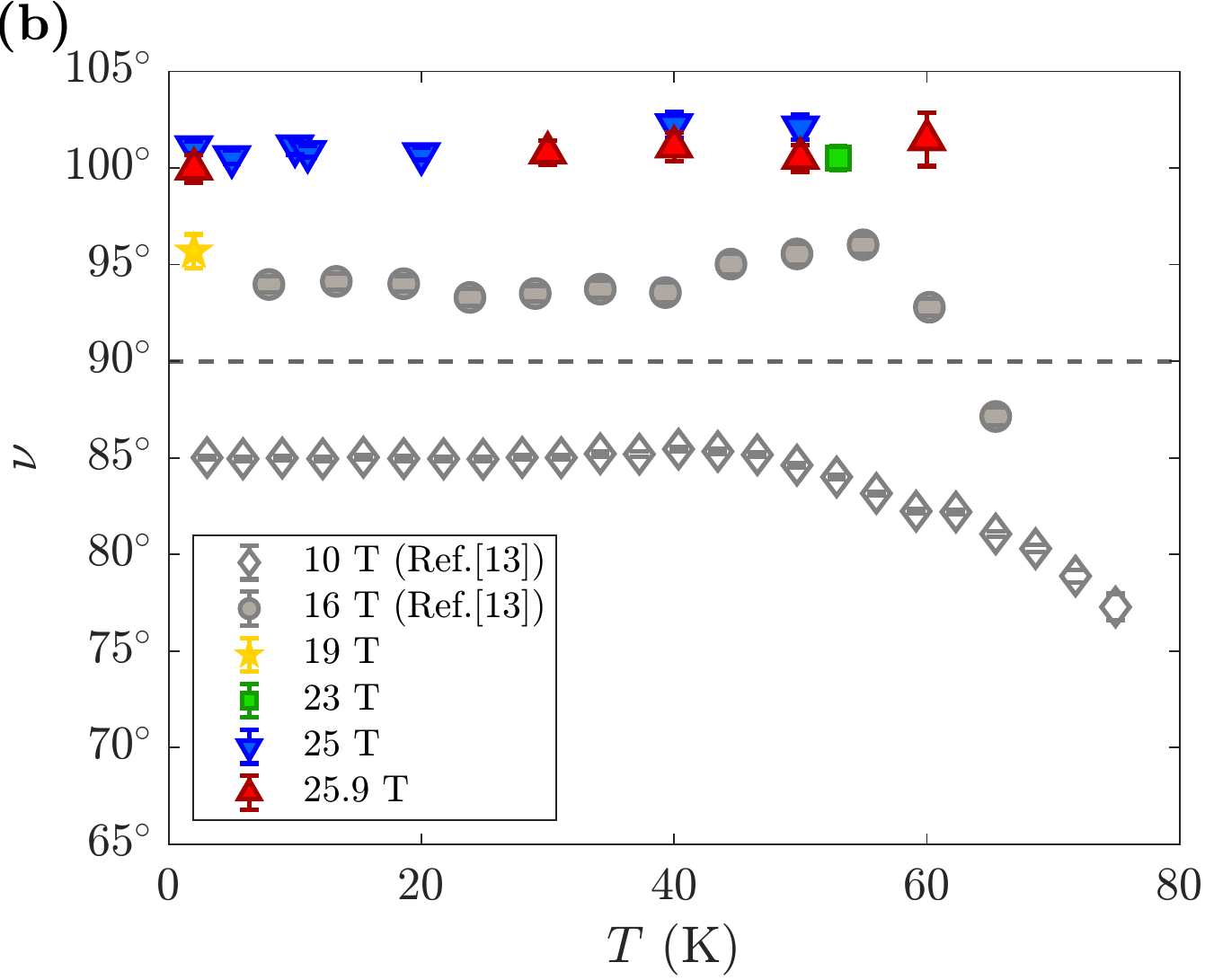}
\caption{(a) The evolution of the vortex lattice structure with magnetic field at 3 K. The circular points are from this study while the filled square points are from a previous study \cite{White2011} up to 11 T, and the open squares are included from ref.~\cite{Cam14}. (b) Variation of the opening angle with temperature at 25, 25.9~T with two single points at 19~T and 23~T.  The diamond and circular points are from a previous study \cite{Cam14} and are provided for reference.}
\end{figure}


The VL form factor is the magnitude of a Fourier component of the spatial variation of the magnetic field within the VL. The form factor $F({\bf q})$ for a diffraction spot with wavevector {\bf q}, is related to the integrated intensity of a diffraction spot, $I({\bf q})$, by the Christen formula \cite{Chr77}

\begin{equation}\label{first}
 |F(\mathbf{q})|^2 = \frac{\Phi_0^2} {2 \pi V \big(\frac{\gamma}{4}\big)^2}  \times  \frac{q I(\mathbf{q})}{ \phi \lambda_n^2},
\end{equation}
where $\Phi_0$ (= $h/2e$) is the magnetic flux quantum, $V$ is the illuminated sample volume, $\gamma$ (= 1.91) is the neutron magnetic moment in nuclear magnetons, $\phi$ is the incident neutron flux per unit area in the neutron wavelength range $\Delta \lambda_n$ centred on $\lambda_n$. $qI(\mathbf{q})$ is derived from integrals over $q_z$ of the neutron counts, $I(q_x, q_y, q_z, \lambda_n) \Delta \lambda$ which arrive in pixels of {\bf q}-space centered on $(q_x, q_y, q_z)$.  After background subtraction, the total integrated intensity is obtained from the region of {\bf q}-space containing a single diffraction spot and from the entire spectrum of neutron wavelengths used, which give a range of $q_z$. The VL peak width in $q_z$ may be obtained by fitting a Gaussian line shape to the intensity summed over $(q_x, q_y)$ as a function of $q_z$.

The dependence of the form factor on field and temperature is shown in Fig.~3. These results are discussed in more detail in the next section, along with the fits included in Fig.~3.

\section{Discussion}

To understand the evolution with field of the VL structure at base temperature, we must consider the whole field range that has been explored in fully oxygenated YBCO. Firstly, at the lowest fields below $\sim 2$~T, the VL structure is distorted hexagonal \cite{White2011,Bro04,White2009}. The distortion of $\sim 30 \%$ is independent of field and its sign indicates an enhanced superfluid density along the $\bf{b}$-direction, which no doubt arises from the superconductivity of the carriers in the CuO chains, which run along this direction. This may be described by anisotropic London theory \cite{Kog81}, which applies when values of the London penetration depth $\lambda_L$ and the vortex spacing are both much larger than the vortex core diameter $\sim \xi$, the coherence length.  
The stronger superconductivity along $\bf{b}$ is confirmed by zero-field measurements of the angles of the nodes in the order parameter \cite{Kir06}. In a purely $d$-wave superconductor, these would lie at exactly $45^{\circ}$ to both $\bf{a}$ and $\bf{b}$ axes, whereas they are found to be closer to $\bf{a}$ ($\sim 40^{\circ}$). As represented schematically in Fig.~4, this indicates enhanced superconductivity along $\bf{b}$. 

\begin{figure}[h!]
\includegraphics[width=\columnwidth]{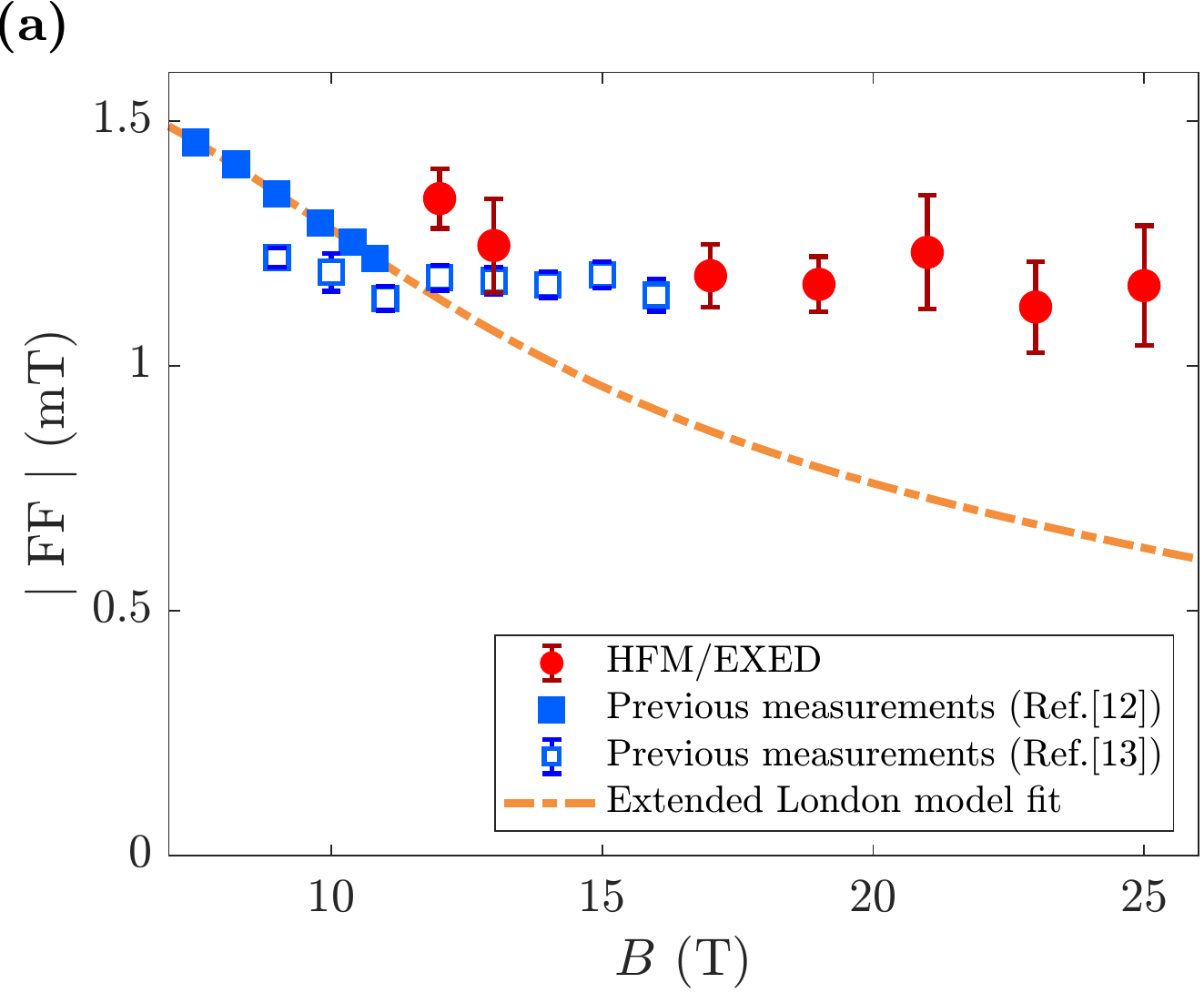}
\includegraphics[width=\columnwidth]{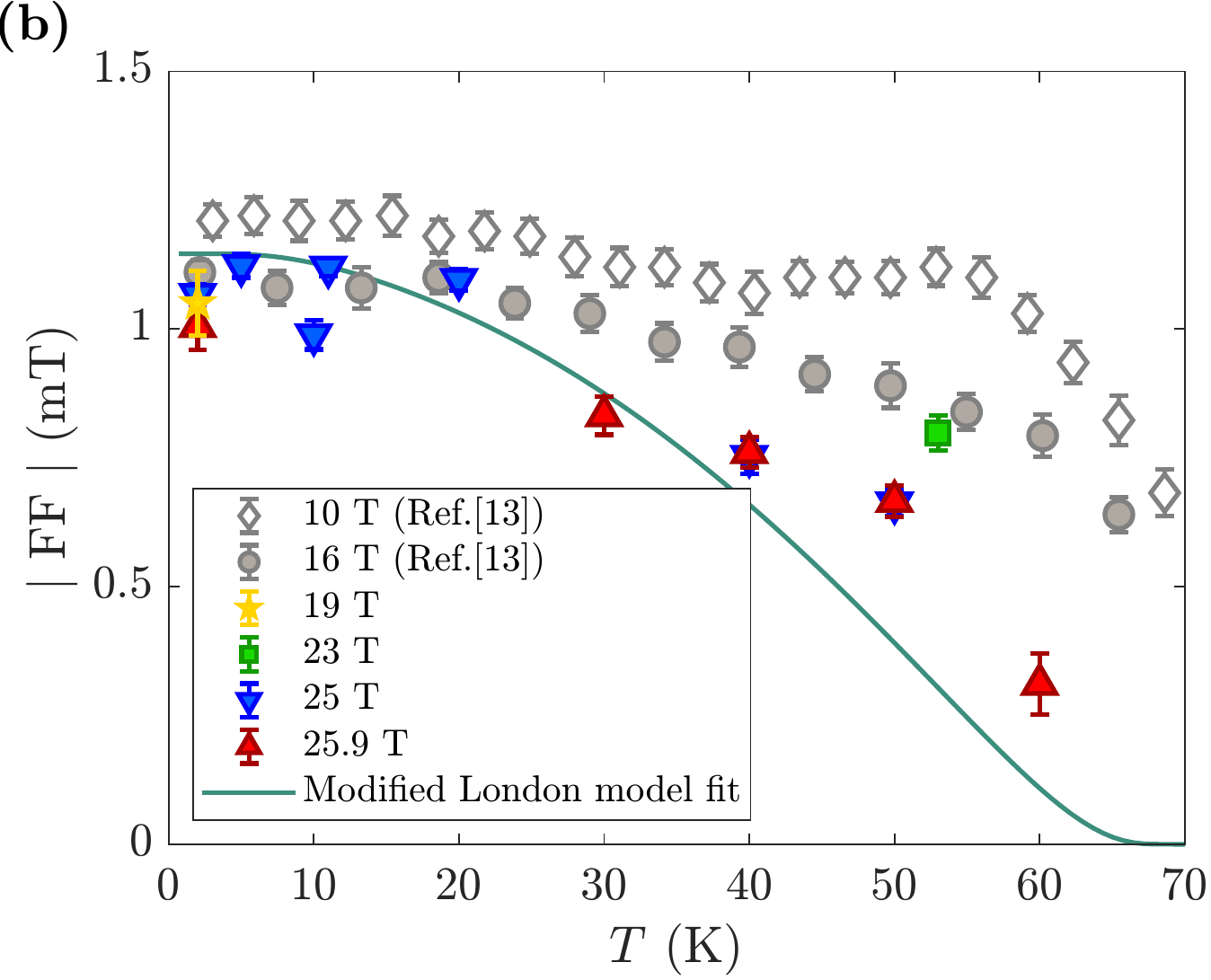}
\caption{(a) The vortex lattice form factor as a function of magnetic field.  The circular points are the new data reported here; the square points are drawn from Refs.~\cite{White2011,Cam14}.  The dashed line is a fit to the filled square points using an extended London model specified in the main text.  (b) The temperature dependence of the vortex lattice form factor.  The triangle and square points are from this work and the diamond and circular points are included from Ref.~\cite{Cam14} for comparison.  The solid line is discussed in the main text. Our data in this panel were taken in the second experiment, and the values of form factor have been multiplied by a normalization constant of 1.5 to account for a loss in intensity observed due to a change in the detector between our first and second experiments.  This normalization constant has been confirmed by a third experiment performed at the same facility with a 15\% Ca-doped YBCO sample.
}
\end{figure}

Between $\sim 2$ and $6.5$~T, the VL is also distorted hexagonal, but with the hexagon rotated by $90^{\circ}$ relative to the low-field case \cite{White2011,Bro04,White2009}.  The distortion falls with increasing field, and this behavior has been discussed elsewhere~\cite{White2011}. 
 For our present purposes, we merely emphasize the reduction of anisotropy with field, which strongly suggests that the chain carrier superconductivity is suppressed by increasing field to a much greater extent than that of the CuO$_{\rm 2}$ planes. 

Finally, above  $\sim 6.5$~T, the VL adopts a high-field centered-rectangular arrangement~\cite{White2011,Bro04,White2009}, continuously connected to what is observed in our field range. We note that the nearest-neighbor vortex pattern is exactly the same as the pattern of diffraction spots around the main beam - but rotated by $90^{\circ}$ about the field axis (this simple relationship represents the transformation between real and reciprocal lattices for the 2-dimensional VL). From Fig.~2~(a), at approximately 11.5~T, the centered rectangle passes through a square arrangement. At this field, the nearest-neighbor vortex directions are at $45^{\circ}$ to the $\bf{a}$ and $\bf{b}$ axes, and change by less than $\pm5^{\circ}$ from this value ($\nu$ is between $80^{\circ}$ and $100^{\circ})$ over the whole field range in which this VL structure is observed. This strongly suggests that the VL arrangement is connected to the nodes in the order parameter, which would be at $45^{\circ}$ if YBCO were a pure $d$-wave superconductor. Strong support for this idea is provided by calculations using first-principles Eilenberger theory~\cite{Ich99} applied to a $d$-wave superconductor. These predict a first order transition from a low-field hexagonal to a high-field square VL, with the VL nearest-neighbor directions at high fields along the nodes of the $d$-wave order parameter.

If we assume that in YBCO above $\sim$6.5~T the VL nearest-neighbor directions are closely linked to the nodal directions, then the variation of the VL structure with field shown in Fig.~2~(a) may be interpreted as an indication of the movement of the nodal directions. Firstly, at fields $\sim 7$~T, we deduce that the nodes are closer to the $\bf{a}$-direction than $\bf{b}$, which is consistent with the superconductivity being stronger along $\bf{b}$ than $\bf{a}$. This is clearly consistent in sign with the anisotropy in $\lambda_L$ shown by the VL at low fields~\cite{White2011} and the direct measurement of nodal positions at zero field~\cite{Kir06}. However as we have seen, the $\bf{b}$-direction superconductivity is weakened by field, and this trend is expected to continue in the high-field region. This behavior is confirmed by the progressive movement of the VL structure towards square at $\sim 11.5$~T. From the continuation of this trend at higher fields \emph{past} the square configuration, we deduce that in this region superconductivity is stronger along $\bf{a}$, giving nodal directions closer to $\bf{b}$, as indicated in Fig.~4. This suggests that the superconductivity in the chain carriers is sufficiently weakened by field that they tend to de-pair the plane carriers also.

We recognise that the decomposition of the carriers into chain and plane is a simplification, since they hybridize where the energy bands cross. Also, the electronic structure of the plane carriers is not quite the same along $\bf{a}$ and $\bf{b}$, so there will be orthorhombic basal plane anisotropies, which may pull the VL nearest-neighbors slightly away from the nodal directions. Nonetheless, the variation of the VL structure with field shows that the basal plane anisotropy is field-dependent. It is far more likely that this is due to a field effect on the superconductivity, as we have described, rather than on the underlying electronic band structure. We emphasize that at high fields, the VL diffraction pattern may be described as distorted square with the stretching along  the {\bf a}$^{\star}$ direction \emph{increasing} with field and reaching a maximum anisotropy at $\nu = 100^{\circ}$ (see Fig.~2~(a)). On the other hand, at low fields, the diffraction pattern may be described as distorted hexagonal, with the stretching along  the {\bf a}$^{\star}$ direction \emph{decreasing} with field. At low fields, the change of VL structure with field is clearly due to the field-dependence of the anisotropy of the penetration depth. It is clear that in the high field region, the VL distortion arises from a different mechanism, and our results strongly suggest that this is the change in the positions of the nodes in the order parameter. Nevertheless, the distortion of the VL at \emph{both} low and high fields may be understood as a consequence of the \emph{same} phenomenon: the weakening of {\bf b}-axis superconductivity with increasing field. 

\begin{figure}[h]
\includegraphics[width=\columnwidth]
{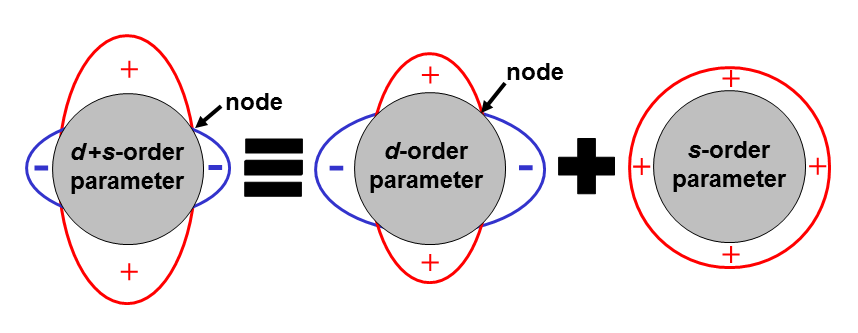}
\caption{Schematic representation of the variation of a $d$ + $s$ superconducting order parameter around a cylindrical Fermi surface. In YBCO, such an admixture must arise because of the crystal structure  and it has the same orthorhombic symmetry. It is seen that the nodal positions lie nearer the direction of weaker superconductivity. This will be the {\bf a}-direction at low fields and the {\bf b}-direction at high fields.}
\end{figure}

We found surprising results on observing the evolution with temperature of the vortex lattice structure at 25~T and 25.9~T in Fig.~2~(b). It has been predicted that the vortex lattice will return to the hexagonal arrangement on approaching $T_{\rm c}$. However, we observe that the VL angle $\nu$ is frozen at $100^{\circ}$ and there is no evidence of a decrease, even at 60~K. During this experiment, the beam did not have enough intensity to measure the vortex lattice closer to its melting point. This may suggest that the change on the vortex lattice structure could happen even closer to $T_{\rm c}$, which is in good agreement with what have been observed in previous SANS studies~\cite{White2011,Cam14}. It is important to point out that we did not observe any variation between 40~K and 60~K, unlike previous observations at 16~T.

Now we turn to the field- and temperature-dependence of the form factor, shown in Fig.~3(a) and (b). For fields much less than $B_{\rm c2}$, it is expected that the form factor will obey a London model, extended to include the effects of overlapping vortex cores of size $\sim \xi$, the coherence length. The key result of this work is that the observed field dependence cannot be fitted by the extended London model, with the VL remaining more robust at high fields than would be expected from the suppression of spatial Fourier components of the field by VL core overlap. 

To illustrate this, we have included in Fig.~3(a) a fit to an extended London model that represents our best attempt with this type of model. We find that the results of first-principles numerical calculations~\cite{Ich99,Ich07} of the form factor at $T << T_{\rm c}$  are not in agreement with Ginzburg-Landau theory (which is only strictly valid close to $T_{\rm c}$). Instead, they are closely approximated by an exponential factor \cite{White2011}:
\begin{equation}\label{second}
F(\textbf{q})  =  F_{\rm London}(\textbf{q})\times \textrm{exp}(- c q^2 \xi^2 ).
\end{equation}
Here, $c$ is a constant that is predicted to lie between 1/4 to 2. In agreement with previous practice \cite{White2011, Cam14}, we chose $c$ to be $\sim 0.44$ (see the Appendix for a detailed justification for this choice). In Eq.~2, we have ignored any $a$-$b$ anisotropy in $\xi$ because throughout our field range, \textbf{q} remains approximately equidistant in angle from both axes. However, we have to take account of the anisotropy of the London penetration depth, because, by assumption, the degree of superconducting pairing, and hence one of the penetration depths in the basal plane, is field-dependent. We therefore introduce values $\lambda_{a}$ and $\lambda_{b}$, arising from supercurrents along the {\bf a} and {\bf b} directions, so that the London equation for the form factor becomes anisotropic
\begin{eqnarray}\label{third}
F\left(q\right)=\frac{\langle B\rangle}{1+q^{2}\lambda^{2}}\rightarrow
\frac{\langle B\rangle}{1+q_{x}^{2}\lambda_{b}^{2}+q_{y}^{2}\lambda_{a}^{2}}.
\end{eqnarray}
We have proposed that the value $\lambda_{b}$ for the chain-direction currents is field-dependent, as the chains become depaired. This happens over a field range around 10~T, so we take for this variation a purely phenomenological expression that has the expected qualitative behaviour of flattening out at large and small fields,
\begin{equation}\label{fourth}
\lambda_{b}^{2}(B)  =  \lambda_{a}^{2}\lbrace 1 + 0.4 \cdot \tanh\left[ (B-10)/7 \right] \rbrace.
\end{equation}
Here, $B$ is in Tesla, and we take the approximate width of the field range where $\lambda_{b}$ is varying as 7~T. The factor 0.4 means that the two penetration depths differ by $\pm \sim 20$\% at low and high fields, with $\lambda_{b}$ shorter than $\lambda_{a}$ at low field~\cite{White2011} and longer at high field.

To calculate the form factor as a function of field, we need the values of $q_x$, $q_y$ and $q$, which may be obtained from the positions of the diffraction spots. Alternatively, using only the value of $B$, the experimentally-determined value of $\nu$, and the fact that each vortex contains one flux quantum, one may write:
\begin{equation}\label{fifth}
q^2  =  4\pi^2 B /\Phi_0 \sin(\nu)~ ; ~(q_x, q_y) = q( \sin(\nu/2), \cos(\nu/2)).
\end{equation}
The exponential in Eq.~\ref{second} for the form factor relies on the value of $\xi$, which may be related to the upper critical field using the Ginzburg-Landau relationship $B_{\rm c2} = \Phi_0 / 2\pi\xi^2 $. Hence, the experimental value of $B_{\rm c2}$ may be used to give the expected value for $\xi$. Alternatively, by substituting for $\xi$ in Eq.~\ref{second}, we may show that the cores give an approximately exponential falloff with field: 
\begin{equation}\label{sixth}
\exp( - c q^2\xi^2) = \exp( -2\pi c  B / B_{\rm c2} \sin(\nu)). 
\end{equation}
In Fig.~3(a), the dashed line represents the field-dependence of $F$ given by Eqs. \ref{second},\ref{third} and \ref{fourth} from a fit to the data at 11 T and below, which gave $\lambda_a = 172.0(8)$~nm and $B_{\rm c2} = 85(3)$ T.   This fails to describe the high field data, and in addition the value of $B_{\rm c2}$ is significantly lower than the reported value of around 120 T \cite{Sekitani2007}.  The falloff at high fields is much slower than that expected from the model and cannot be reproduced by the extended London model expressed in Eqs.~\ref{second}, \ref{third} and \ref{fourth}. 

The VL perfection revealed by the present data (see Appendix) is no better than that observed at low fields. Hence any effects of VL pinning in our field range should mimic a $\xi$ with a value larger than that calculated from $B_{\rm c2}$. However, Fig.~3(a) shows that the form factor is not falling off with field as expected for any reasonable value of $\xi$ and a constant value of $\lambda_L$. Furthermore, if we assume that the field is leading to the weakening of superconductivity along the CuO chain direction, which we represent using the field-dependent $\lambda_b$ in Eq.~\ref{fourth}, we can use a reasonable value of $\xi$ (corresponding to $B_{c2} \sim 85$~T) to fit the data up to  $\sim11$~T but not to higher fields.

Fig.~3(b) shows the variation of the form factor with temperature at different fields. The value at high temperatures shows a clear decrease with field. We have fitted our temperature dependence assuming  $d+s$ pairing \cite{Pro06}, where
\begin{equation}
    \Delta (T, \varphi) = \Delta_{0,d} (T) \cos(2 \varphi) + \Delta_{0,s} (T)
\end{equation}
where $\Delta_{0,s} (T) = -\cos(100^{\circ}) \Delta_{0,d} (T)$ to give nodes at the observed angle. We also assume that $\Delta_{0,d} (0) = 2.14 k_{\rm B} T_{\rm c}$ \cite{Pro06}, with $B_{c2} = 120$~T \cite{Sekitani2007} and $T_{\rm c} = 70$~K at 25~T \cite{Grissonnanche2014}. This model also fails to follow the temperature dependence of the form factor at high fields, especially at temperatures above 40~K, indicating again that the expected description breaks down.

The intensity of the VL signal reflects the field contrast between the cores and their surroundings. We therefore conclude that, at high fields (although low relative to $B_{\rm c2}$), there is a contribution to the spatial variation of magnetic field in the VL in addition to that arising from super-currents circulating around the vortices. This extra contribution must correspond to an additional magnetization of the vortex cores. This can arise as follows: the quasi-particles in the VL cores do not have to adopt the anti-parallel spin arrangement of Cooper pairs, so the spins may align parallel to the magnetic field. This allows the formation of a Pauli paramagnetic moment in the core region~\cite{Ich07}. Such effects must be present in all singlet-pairing superconductors, but will be negligible unless $\mu_{\rm B} B_{\rm c2} \geq k_{\rm B} T_{\rm c}$, so that the Zeeman energy of the electron spins is comparable with the zero-field energy gap. Pauli paramagnetic effects have been observed in heavy-fermion materials such as CeCoIn$_5$ \cite{Bia08,Whi10}, CeCu$_2$Si$_2$ \cite{Campillo2021}, a borocarbide \cite{DeB07} and an iron-based superconductor \cite{Kuhn16}, but not to our knowledge in a high-$T_{\rm c}$ cuprate. Nevertheless, Pauli-paramagnetic effects are expected in our sample, because it satisfies $\mu_{\rm B} B_{\rm c2} \simeq k_{\rm B} T_{\rm c}$.

\section{\label{sec:level1}Conclusions}

Using neutron scattering at a unique instrument, we have observed diffraction by the lattice of magnetic flux vortices in a superconductor at higher fields than ever before. Our results for YBa$_2$Cu$_3$O$_7$ are a clear indication that high magnetic fields tend to destroy superconducting pairing in the carriers traveling along the crystal \textbf{b} direction (CuO chains) in this material. This leads to a field-dependent change in the superconducting anisotropy, which will be reflected in a change in the angular position of the order-parameter nodes in this orthorhombic $(d + s)$-wave material. In addition, we find that the intensity of the diffraction signal from the vortex lattice hardly falls off at high fields and the standard models do not account for the field and temperature dependencies of the form factor for these fields. We take this is an indication of Pauli-paramagnetic vortex cores. Our results bode well for further studies at the high-field frontier when still greater steady fields become available at neutron scattering facilities.

EMF was supported by the Leverhulme Foundation. ASC acknowledges support from the German Research Foundation (DFG) under Grant No. IN 209/3-1. We would like to thank Robert Wahle, Sebastian Gerischer, Stephan Kempfer, Peter Heller, Klaus Kiefer and Peter Smeibidl for their support during the experiment.

\section{\label{sec:level1}Appendix}

\subsection{Sample}
The sample, prepared at the  Walther Meissner Institut, consisted of a mosaic of eleven co-aligned single crystals of de-twinned YBa$_2$Cu$_3$O$_7$ with a total mass of $\sim 73$ mg. The crystals were grown from a molten flux of BaCO$_{3}$, CuO and Y$_{2}$O$_{3}$ in BaZrO$_{3}$ crucibles \cite{Erb96}. They were de-twinned through the application of uniaxial stress at a temperature of 500~C for $24$ hours \cite{Lin92, Hin07}. The crystals were then oxygenated close to the O$_7$ composition under an O$_{2}$ atmosphere of $100$ bar at $300^{\circ}$ C for $150$ hours \cite{Erb99}. The filled CuO chains made the crystals slightly over-doped, but greatly reduced pinning by oxygen vacancies relative to that for an optimally-doped sample. A crystal from the mosaic gave a zero field \textit{T}$_{c}$ of $89.0$ K by SQUID magnetometry in a field of $1$ mT, with a $90$\% transition width of $2$ K. Given the high purity of the samples, this spread in \textit{T}$_{c}$ suggests a slight spread in oxygen content across the mosaic. The mosaic was mounted on a $1$~mm thick aluminium plate, with the crystal \textbf{c} axis perpendicular to the plate and the \textbf{a} direction co-aligned between crystals. 

\subsection{Theoretical calculations of Vortex Lattice Form Factor}

\begin{figure}[h!]
\includegraphics[width=\columnwidth]{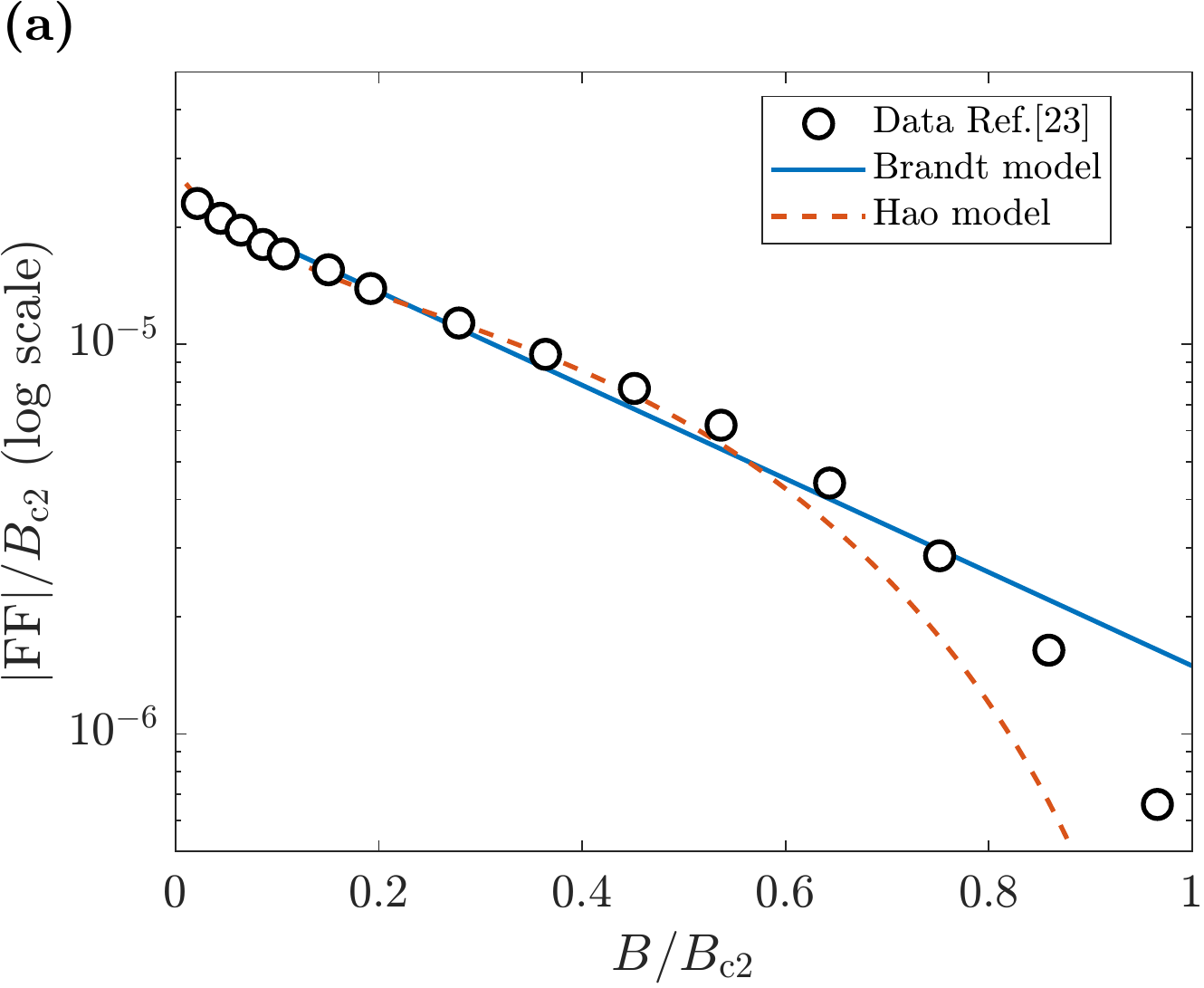}
\includegraphics[width=\columnwidth]{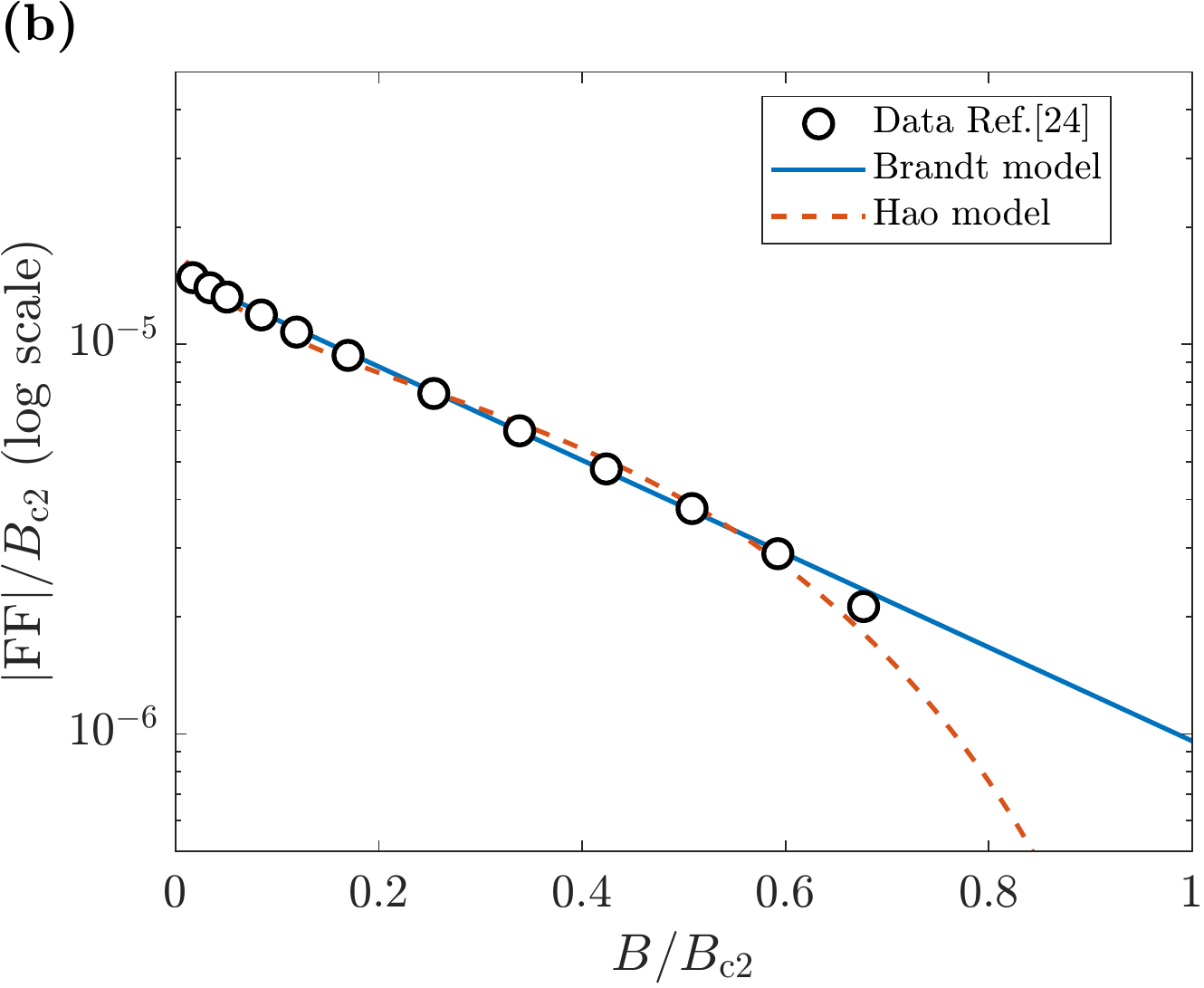}
\caption{Theoretical form factors versus field from numerical solution of the Eilenberger equations (points), the Hao-Clem variational solution of the GL equations (dashed line) and London model with exponential cut-off or Brandt model (solid line) versus field. Theoretical form factors calculated (a) for $T= 0.5T_{\rm c}$ using results from \cite{Ich99} and (b) for $T= 0.1T_{\rm c}$ using results from \cite{Ich07}.}
\end{figure}

There are just two algebraic theories giving the expected form factor in the mixed state at fields not close to $B_{\rm c2}$, but they are both of limited validity. One is London theory, in which vortex cores are ignored, and hence this theory is only valid for very large $\kappa = \lambda/\xi$ (true for YBCO) and for $B << B_{\rm c2}$. The other is Ginzburg-Landau (GL) theory, developed as an expansion in powers of the order parameter near $T_{\rm c}$, and hence only numerically valid in this region. It does however give a \emph{qualitative} picture of the mixed state at lower temperatures. To obtain an explicit expression for the predictions of GL theory, Hao \& Clem carried out a variational solution \cite{Hao91}. However, there is a \emph{numerical} first-principles method for obtaining the mixed state structure, using the Eilenberger equations \cite{Ich99,Ich07}, which can be applied at lower temperatures and be used to test the validity of the GL equations away from $T_{\rm c}$. From Fig.~5, it is clear that there is a better agreement between the Clem-Hao model and the Eilenberger approach at low temperatures rather than higher. The Eilenberger result is very close to a straight line on a log-linear plot for fields $< 0.5 B_{\rm c2}$. A fitting of the low-field region, up to $0.5 B_{\rm c2}$, gives the form:
\begin{equation}\label{number}
F(\textbf{q})  =  F_{\rm London}(\textbf{q})\times \textrm{exp}(- c q^2 \xi^2 ).
\end{equation}
with c = 0.44. As we can observe, we get a better agreement to the model at low temperatures, as shown in Figure~5 (b) for the $T = 0.1 T_c$ case, but even leaving c as a free fitting parameter at the 0.5$T_c$ case we get a similar slope with c = 0.41(2). This London model with exponential cutoff (or Brandt model) was also found to give a better fit to experimental data than the Hao-Clem expression. Further details may be found in Ref.~\cite{Bow08}.

\subsection{Vortex Lattice perfection}

\begin{figure}
\includegraphics[width=\columnwidth]{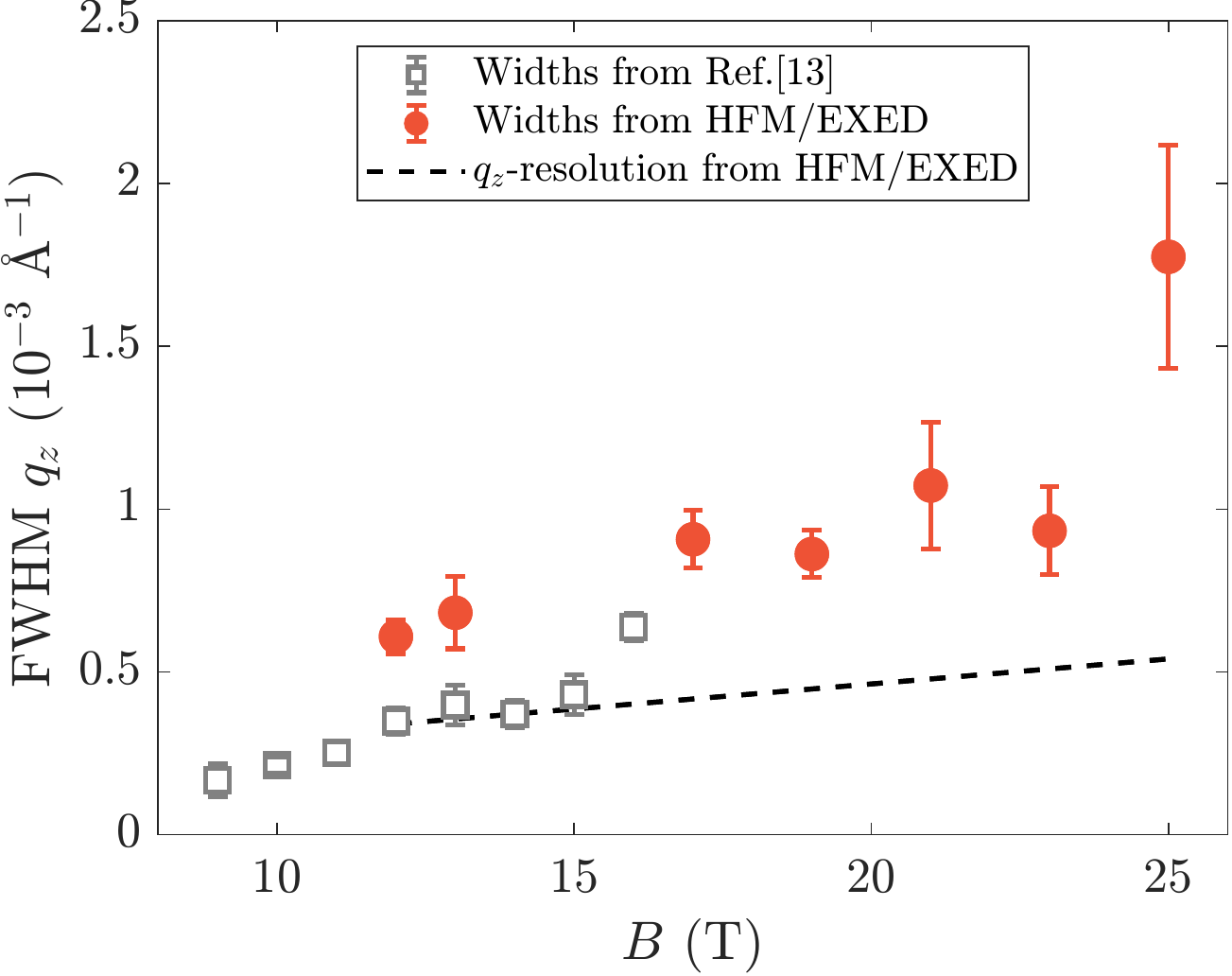}
\caption{The $q_z$ width of the diffraction spots at base temperature versus field above 8 T. The empty square points are from a previous study \cite{Cam14}.}
\end{figure}

In addition to the quantities in the main text, we could also obtain the $q_z$ widths of the diffraction spots, which are shown in Figs.~6.  Above 17 T, the FWHM is around  $q_z$  $\sim  10^{-3}$ \AA$^{-1}$ which is about $2-3 \times$ the instrument resolution. This $q_z$ width corresponds to a correlation length along the vortex lines of $\sim 3 \times 10^{-7}$~m. Above 17 T, the $q_z$ widths seems to increase with field, following the trend given by data from previous studies \cite{Cam14}. The widths of the diffraction spots in the other two directions are largely instrument-limited at a value $\sim 8 \cdot 10^{-3}$ \AA, so they give little information about VL perfection.


%

\end{document}